\documentclass[
reprint,
superscriptaddress,
showpacs,preprintnumbers,
 amsmath,amssymb,
 aps,
prx,
]{revtex4-1}

\usepackage{graphicx}
\usepackage{dcolumn}
\usepackage{bm}
\usepackage{amsmath}
\usepackage{esint}
\PassOptionsToPackage{normalem}{ulem}
\usepackage{ulem}
\usepackage{soul,xcolor}
\usepackage{color}
\usepackage{comment}
\usepackage[utf8]{inputenc}
\DeclareUnicodeCharacter{00A0}{~}

\usepackage{geometry}
\begin{document}

\setstcolor{red}

\definecolor{oradsblue}{HTML}{B8EEFF}
\definecolor{zahirred}{HTML}{FF8080}
\newcommand{\OR}[1]{~{\sethlcolor{oradsblue}\hl{(OR:~#1)}}}  
\newcommand{\ZA}[1]{~{\sethlcolor{zahirred}\hl{(ZA:~#1)}}}


\title{On modifications of fundamental radiative processes in near-zero-index media of various dimensions}

\author{Micha\"el Lobet}
\thanks{These authors contributed equally to this work.}
\affiliation{John A. Paulson School of Engineering and Applied Sciences, Harvard University, 9 Oxford Street, Cambridge, MA 02138, United States of America}
\affiliation{ Centre Spatial de Liège, Avenue du Pré-Aily, B-4031 Angleur, Belgium}

\author{I\~nigo Liberal}
\thanks{These authors contributed equally to this work.}
\affiliation{Electrical and Electronic Engineering Department, Universidad P\'{u}blica de Navarra,
Campus Arrosad\'ia, Pamplona, 31006 Spain}
\author{Erik N. Knall}
\affiliation{John A. Paulson School of Engineering and Applied Sciences, Harvard University, 9 Oxford Street, Cambridge, MA 02138, United States of America}
\author{M. Zahirul Alam}
\address{Institute of Optics, University of Rochester, Rochester, NY, USA}
\address{Department of Physics, University of Ottawa, Ottawa, Ontario, Canada}

\author{Orad Reshef}
\affiliation{Department of Physics, University of Ottawa, Ottawa, Ontario, Canada}
\author{Robert W. Boyd}
\address{Institute of Optics, University of Rochester, Rochester, NY, USA}
\address{Department of Physics, University of Ottawa, Ottawa, Ontario, Canada}
\author{Nader Engheta}
\affiliation{Department of Electrical and Systems Engineering, University of Pennsylvania, Philadelphia, PA 19104, USA} 
\author{Eric Mazur \email{mazur@seas.harvard.edu}}
\affiliation{John A. Paulson School of Engineering and Applied Sciences, Harvard University, 9 Oxford Street, Cambridge, MA 02138, United States of America}
\date{\today}

\begin{abstract}
Spontaneous emission, stimulated emission and absorption are the three fundamental radiative processes describing light-matter interactions. Here, we theoretically study the behaviour of these fundamental processes inside an unbounded medium exhibiting a vanishingly small refractive index, i.e., a near-zero-index (NZI) host medium. We present a generalized framework to study these processes and find that the spatial dimension of the NZI medium has profound effects on the nature of the fundamental radiative processes. Our formalism highlights the role of the number of available optical modes as well as the ability of an emitter to couple to these modes as a function of the dimension and the class of NZI media. We demonstrate that the fundamental radiative processes are inhibited in 3D homogeneous lossless zero index materials but may be strongly enhanced in a zero index medium of reduced dimensionality. Our findings have implications in thermal, nonlinear and quantum optics as well as in designing quantum metamaterials at optical or microwave frequencies.

\end{abstract}

\maketitle



In 1916 and 1917, Einstein proposed three fundamental radiative processes to explain light-matter interactions: spontaneous emission, stimulated emission and absorption \cite{Einstein1916,Einstein1917}. Einstein's $A_{21}$, $B_{21}$ and $B_{12}$ coefficients are typically used to describe the rate of these processes, respectively. Later, Purcell demonstrated that the spontaneous emission rate is not an immutable property of matter and that the environment can significantly modify it \cite{Purcell1946}. In recent times, there are ongoing intensive research efforts in designing nanostructured materials to control the spontaneous emission rates for applications in quantum optics, quantum computing, quantum communications, and quantum chemistry \cite{Yablonovitch1987,Lodahl2015,chikkaraddy2016single}. Most notably, metamaterials, artificial materials that may exhibit electromagnetic (EM) properties otherwise absent in natural materials, have been explored for that purpose due to their ultimate flexibility in tailoring the local optical environment. These engineered materials may feature extreme parameters such as a near-zero refractive index (NZI), and have been shown to exhibit exotic electromagnetic properties~\cite{Liberal2017,Engheta2006,Silveirinha2006,Ziolkowski2004,Vulis2019}. 

As a consequence of a vanishing refractive index at frequency $\omega_Z$, the phase velocity $v_\varphi$ of an EM wave inside a near-zero index material diverges and the wavelength $\lambda$ of the wave is significantly stretched. Since the refractive index is defined as $n(\omega)=\sqrt{\varepsilon(\omega)\mu(\omega)}$, $\varepsilon(\omega)$ the relative permittivity and $\mu(\omega)$ the relative permeability, three different routes exist to achieve an NZI response: $\varepsilon$ approaches zero with arbitrary $\mu$ (i.e., epsilon-near-zero (ENZ) media) \cite{Silveirinha2006,Edwards2008}; $\mu$ approaches zero  with arbitrary $\varepsilon$ (i.e., mu-near-zero (MNZ) media) \cite{Marcos2015}; and both $\varepsilon$ and $\mu$ simultaneously approach zero (i.e., epsilon-and-mu-near-zero (EMNZ) media) \cite{Vulis2019, Ziolkowski2004,Mahmoud2014,Li2015,Briggs2013}. Although all three classes of NZI media share a near-zero refractive index, they differ critically in other characteristics. For example, the normalized wave impedance $Z\left(\omega\right)=\sqrt{\mu\left(\omega\right)/\varepsilon\left(\omega\right)}$, tends to infinity in ENZ media, $Z\left(\omega_Z\right)\rightarrow\infty$, to zero in MNZ media, $Z\left(\omega_Z\right)\rightarrow 0$, and to a finite value in EMNZ media, $Z\left(\omega_Z\right)\rightarrow\left.\sqrt{\partial_{\omega}\mu\left(\omega\right)/\partial_{\omega}\varepsilon\left(\omega\right)}\right|_{\omega\rightarrow\omega_Z}$. 
Similarly, the group index, $n_g\left(\omega\right)=c/v_g\left(\omega\right)$ (where $v_g\left(\omega\right)$ is the group velocity) tends to infinity in ENZ and MNZ unbounded lossless media \cite{Javani2016}, while it has a finite value, $\omega \partial_{\omega}n\left(\omega\right)$, in EMNZ media \cite{Vulis2019, Ziolkowski2004}. 
Consequently, the selected class of NZI medium makes a profound impact on different optical processes, including propagation, scattering and radiation of EM waves~\cite{Liberal2017}.

Similarly, extreme material parameters impact fundamental radiative processes and their associated transition rates. Specifically, complete inhibition of spontaneous emission was predicted for three dimensional (3D) ENZ and EMNZ media \cite{Liberal2016ScAd,Liberal2018}, and two-dimensional (2D) implementations of EMNZ media \cite{Mahmoud2014}.
Typically, the suppression of spontaneous emission is justified due to the depletion of optical modes as the refractive index goes to zero. 
This effect is somewhat analogous to the inhibition of spontaneous emission in photonic nanostructures exhibiting a band-gap \cite{Bykov1972,Yablonovitch1987,Yablonovitch1987,John1990,Joannopoulos2008}. However, it is distinct in that  the propagation of electromagnetic waves is allowed in EMNZ media. In contrast, studies of metallic waveguides near cutoff that effectively behave as one-dimensional (1D) ENZ media reveal that the spontaneous emission rate is enhanced (theoretically diverges) in those systems \cite{Alu2009boosting,Fleury2013,Sokhoyan2013,Li2016}. 

The radical difference in the predicted responses, i.e., inhibition versus enhancement, raises the question of whether these effects relate to details of the structural implementation of NZI media (e.g., microscopic coupling to a dispersive waveguide) or if they are an accurate representation of the true material response of NZI media. The latter would then imply a complex interplay between the class of NZI media (ENZ, MNZ and EMNZ) and the dimensionality of the system (1D, 2D, and 3D).

To the best of our knowledge, there is no unified framework encompassing studies of all the fundamental radiative processes for all NZI media classes (ENZ, MNZ and EMNZ) and dimensionalities (1D,2D and 3D). Here, we address this question by presenting a unified framework that provides compact expressions for the transition rates in dimension-dependent NZI media. Our results are relevant for recent experimental demonstrations of various classes of NZI media \cite{Vesseur2013,Briggs2013,Liberal2017photonic,Reshef2017direct,Luo2018coherent}.

To begin, we consider a two-level system, $\left\{\left|e\right\rbrace,\left|g\right\rbrace\right\}$ with transition dipole moment $\mathbf{p}=p\,\mathbf{u}_z $ embedded in a 3D unbounded lossless homogeneous material with a transition frequency $\omega$. First, we evaluate the influence of an NZI background on spontaneous emission, and then we discuss how these conclusions apply to the absorption and stimulated emission processes. To this end,  we follow the macroscopic QED formalism \cite{Vogel2006} so that the Einstein coefficient $A_{21}'$, representing the spontaneous emission rate, can be written as a function of the Green's function $\bm{G}$ as follows \cite{Dung2003}:
\begin{align}
A_{21}'(\omega) &= \frac{2\omega^2}{\hbar \varepsilon_0 c^2}\,\,|\mathbf{p}|^2\, \mathbf{u}_z \cdot {\rm Im}[ \mathbf{G}(\mathbf{r}_0,\mathbf{r}_0,\omega)]\cdot \mathbf{u}_z \nonumber \\
&= \mathrm{Re} \left [ \mu(\omega) n(\omega) \right ] A_{21},
\label{Inhibition3DSpontEm}
\end{align}
where we have used
\begin{equation}
\mathbf{u}_z\cdot{\rm Im}[\mathbf{G}(\mathbf{r}_0,\mathbf{r}_0,\omega)]\cdot\mathbf{u}_z=\frac{\omega}{6\pi c} \mathrm{Re}\left[\mu(\omega)n(\omega)\right]
\end{equation}
for homogeneous media \cite{Dung2003} and $A_{21}=\omega^3|\mathbf{p}|^2/(3\pi\varepsilon_0\hslash c^3)$ is the free-space spontaneous emission coefficient.

\begin{figure*}
\includegraphics[width=\textwidth]{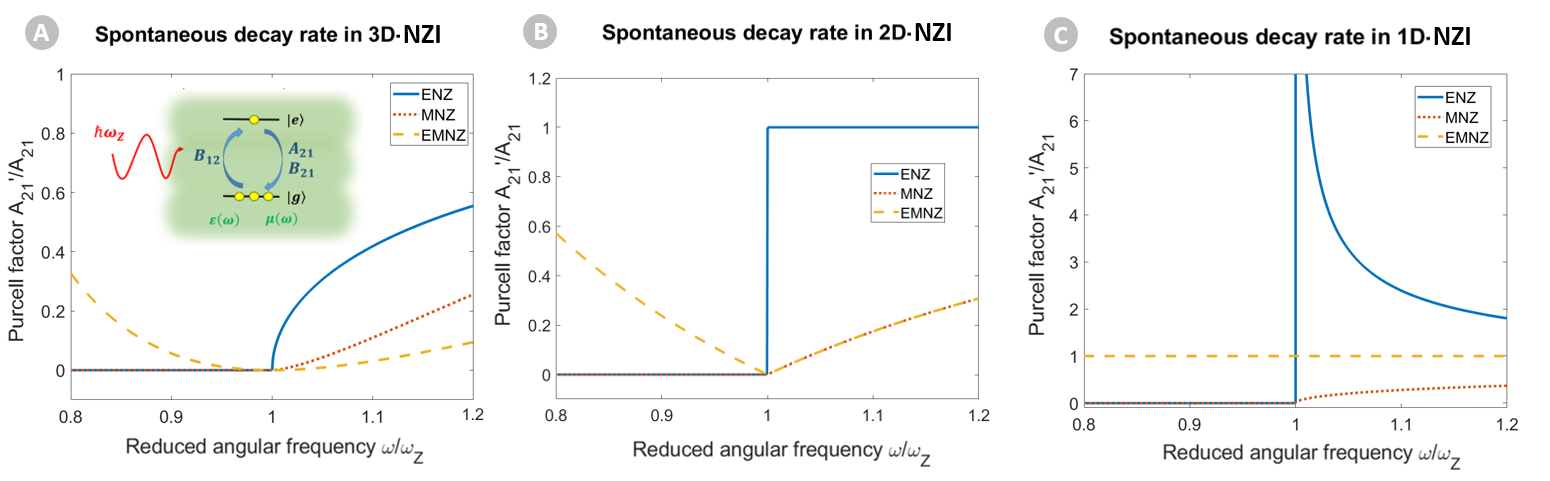}
\caption{Spontaneous decay rate normalized with free-space (Purcell factor) for (a) 3D (b) 2D (c) 1D homogeneous dispersive NZI media. EMNZ metamaterial with a Lorentz model ($\varepsilon(\omega)=\mu(\omega)=\frac{\omega^2-\omega_Z^2+2i\omega\Gamma}{\omega^2-\omega_r^2+2i\omega\Gamma}$, $\omega_r=0.1\omega_Z$, $\Gamma=0$ for lossless case \cite{Supp1}) (yellow), ENZ material with $\varepsilon(\omega)$ and $\mu=1$ (blue), MNZ material with $\mu(\omega)$ and $\varepsilon=2.25$(red). We choose $\omega_\mathrm{max}=\alpha k_B T/\hbar=\omega_Z$ where $\alpha=2.821439$ is a constant and $T=300K$. Inset of (a): two-level system $\left\{\left|e\right\rbrace,\left|g\right\rbrace\right\}$ embedded inside an unbounded, lossless and homogeneous dispersive material.}
\label{fig:Fig1SpontEM3D}
\end{figure*}

We directly conclude from Eq.\,(\ref{Inhibition3DSpontEm}) that spontaneous emission is inhibited in all classes of unbounded lossless 3D NZI media as $n(\omega_Z)\rightarrow 0$.
Fig. \ref{fig:Fig1SpontEM3D}a shows the inhibition of spontaneous emission for the three classes and their different behaviours around the NZI frequency $\omega_Z$ .

Next, we study stimulated emission and absorption by referring to the detailed balance equation \cite{Einstein1916,Einstein1917}. In our case, detailed balance means that spontaneous emission and stimulated emission are balanced by the absorption process. This principle leads to the Einstein relations for dispersive materials \cite{Loudon2000}:
\begin{equation}
\frac{A'_{21}}{B'_{21}} = \mathrm{DOS} \times  \hbar \omega, \,\,\,\,\,\,\,\,\,
\frac{B'_{21}}{B'_{12}} = \frac{g_1}{g_2},
\label{eq:Einstein relations}
\end{equation}


\noindent where $DOS=n^2(\omega)\omega^2/\pi^2c^2v_g(\omega)$ is the density of states and $g_i$ the degeneracy of state $| i \big \rangle$. From here, one can derive a general expression for the Einstein $B'_{21}$ coefficient in a dispersive material:
\begin{align}
B_{21}'(\omega) &=\frac{2\pi^2 c}{\hbar^2 \omega \varepsilon_0 n^2(\omega)n_g(\omega)}|\mathbf{p}|^2 \mathbf{u}_z \cdot {\rm Im}[   \mathbf{G}(\mathbf{r}_0,\mathbf{r}_0,\omega)]\cdot \mathbf{u}_z \nonumber\\
&= \frac{Z(\omega)}{n_g(\omega)}B_{21}.
\label{Milonni2003B}
\end{align}

\noindent Using this formulation, we can evaluate the Einstein $B'_{21}$ coefficient for the different classes of NZI media:
\begin{equation}
  B_{21}'(\omega=\omega_Z) = B_{21}\times
    \begin{cases}
      \frac{1}{n_g(\omega)}\sqrt{\frac{\frac{d\mu(\omega)}{d\omega }\Bigr|_{\large{\substack{\omega=\omega_Z}}}}{\frac{d\varepsilon(\omega)}{d\omega }\Bigr|_{\large{\substack{\omega=\omega_Z}}}}} & \text{for EMNZ materials}\\
      \frac{2}{\omega_Z \frac{d\varepsilon(\omega)}{d\omega}} & \text{for ENZ materials}\\
      0 & \text{for MNZ materials}
    \end{cases}.       
    \label{Eq:B21prim}
\end{equation}

Equations\,(\ref{Milonni2003B}) and (\ref{Eq:B21prim}) show that the $B_{21}$ coefficient is modified by the background medium, as pointed out in previous works \cite{Milonni1995,Milonni2003}. This result suggests that the ratio between spontaneous and stimulated emission can be selected by changing the background material. However, one must be careful to point out that the stimulated emission rate is given by the product of $B_{21}$ and the spectral density of states $\rho(\omega,T)$ ($\Gamma_{\rm sti}=B_{21}\rho(\omega,T)$). In addition, in view of Eq.\,(\ref{eq:Einstein relations}), the absorption rate must be equal to the stimulated emission rate  $\Gamma_{\rm sti}=\Gamma_{\rm abs}$ \cite{Supp1}.

Therefore, in order to elucidate the impact of NZI media on the total stimulated emission and absorption rates,  we address how the spectral energy density of thermal radiation $\rho\left(\omega,T\right)$ behaves in the NZI limit. This procedure will also allow us to study thermal equilibrium radiation for a black-body at temperature $T$ immersed in NZI media. The spectral energy density of thermal radiation in a material is given by \cite{Milonni2003}
\begin{equation}
\rho(\omega,T)=\frac{\hbar\omega^3}{\pi^2 c^3}\,\frac{1}{e^{\frac{\hbar\omega}{k_BT}}-1}\,\,n^2(\omega)\,n_g(\omega).
\label{Eq:Planck}
\end{equation}

\begin{figure}
\includegraphics[width=\columnwidth]{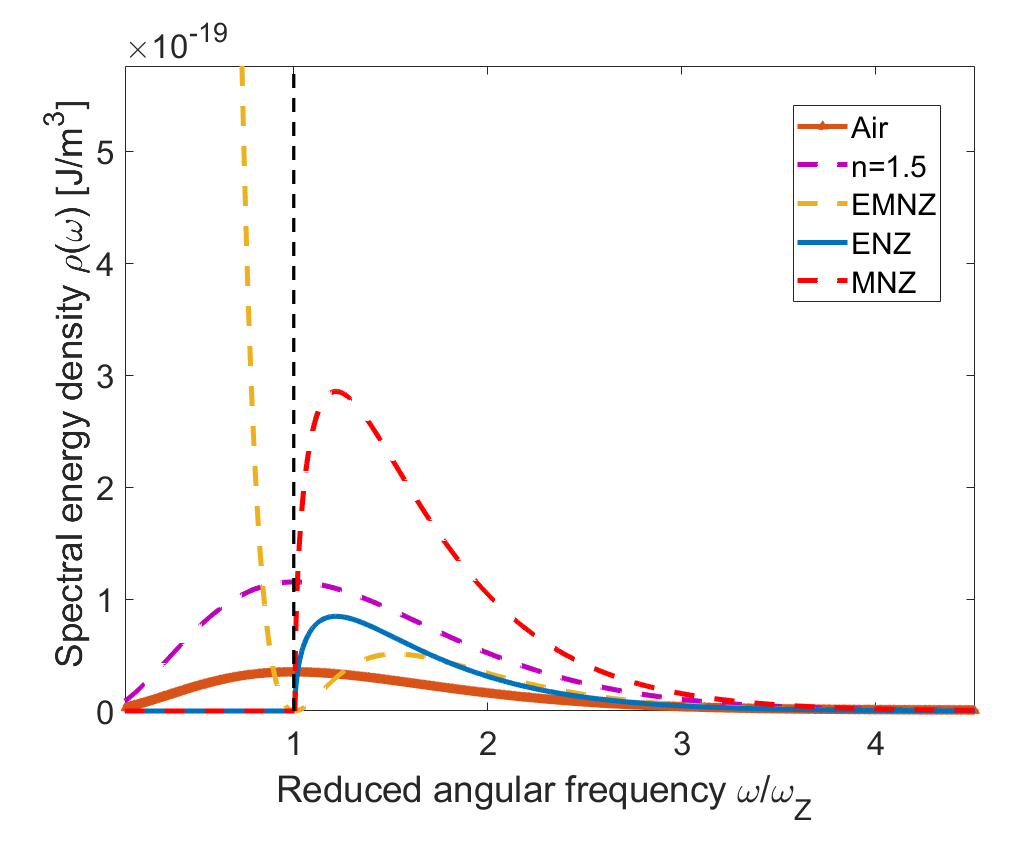}
\caption{Spectral energy density $\rho(\omega$) for air (brown), non-dispersive electric permittivity $\varepsilon=2.25$ (purple), EMNZ metamaterial with a Lorentz model ($\varepsilon(\omega)=\mu(\omega)=\frac{\omega^2-\omega_Z^2+2i\omega\Gamma}{\omega^2-\omega_r^2+2i\omega\Gamma}$ and $\omega_r=0.1\omega_Z$, $\Gamma=0$ for lossless case \cite{Supp1}) (yellow), ENZ material with $\varepsilon(\omega)$ and $\mu=1$ (blue), MNZ material with $\mu(\omega)$ and $\varepsilon=2.25$ (red).Temperature is set to $T=300K$.}
\label{fig:fig1}
\end{figure}

Figure~\ref{fig:fig1} represents the spectral energy density (corresponding to black-body radiation) for the different classes of NZI media. We set the zero-index frequency $\omega_Z$ to be equal to the maximum frequency of the spectral energy density in vacuum $\omega_\textrm{max}=\alpha k_B T/\hbar$ where $\alpha=2.821439$ \cite{Kittel1980}. In EMNZ materials, the group index is constant but the spectral energy density is highly reduced in the NZI spectral region. For frequencies below $\omega_Z$, the allowed propagation corresponds to propagation inside a left handed materials with a refractive index close to zero \cite{Dung2003,Milonni2003}. For ENZ and MNZ media, no propagation is allowed for frequencies below $\omega_Z$ because of the imaginary refractive index.  In general, since $\rho(\omega,T)$ scales as $n^2(\omega)\,n_g(\omega)$, it is reduced in the NZI spectral region and vanishes exactly at $\omega_Z$.
This effect can be intuitively explained by using a box quantization treatment. The spectral energy density of thermal radiation $\rho(\omega)$ is the product of the density of states (DOS) by the mean energy of a state at temperature $T$, $\theta(\omega,T)=\frac{\hbar \omega}{e^{\frac{\hbar\omega}{k_BT}}-1}$ \cite{Joulain2005}. The modes in a 3D box of volume $L^3$ in $k$ space are separated by $\Delta k = \pi/L$.  Consequently, the number of modes in a spherical shell between $k$ and $k+dk$ is 
$\pi k^2dk\left(\pi/L\right)^{-3}$, so that the density of modes scales as $n^2(\omega)\,n_g(\omega)$ \cite{Loudon2000}. When the index is near zero, the number of modes within the sphere is much lower than that in vacuum and the DOS reaches its minimum value. Therefore, the spectral energy density $\rho(\omega,T)$ is equal to zero at the NZI frequency $\omega_Z$ and, consequently, thermal radiation from a black-body immersed in such media would be inhibited.

In addition, by combining Eqs.\,(\ref{Eq:B21prim}) and (\ref{Eq:Planck}) we find that the stimulated emission rate vanishes. Therefore, although previous works pointed out the possibility of controlling stimulated emission \cite{Milonni1995,Milonni2003}, we conclude that all fundamental radiative processes are inhibited inside unbounded 3D homogeneous lossless NZI media, which can be understood as a consequence of the depletion of optical modes around the NZI frequency. The same conclusion can be obtained by directly evaluating the stimulated emission and absorption rates by means of Fermi's golden rule \cite{Supp1}, without needing to invoke the detailed balance equation or thermal equilibrium considerations.
Furthermore, we note that the inclusion of the local-field correction factors using a real-cavity model \cite{Dung2003} does not change the above conclusions (Supplemental Materials (SM) \cite{Supp1}). Moreover, including material absorption gives rise to a finite value of $A'_{21}$, directly proportional to $\mathrm{Im}[ \varepsilon]$, which can be very small \cite{Supp1}.

One has to be careful, however, in translating directly this result to systems with a lower dimensionality. It is worth noticing that the macroscopic QED formalism \cite{Dung2003,Vogel2006} used above provides a very convenient theoretical framework to evaluate radiative transitions based on the imaginary part of the dyadic Green's function. This compact formulation fails however to provide a physical insight on how different classes of NZI media affect radiative transitions. 

To address these issues, we introduce a simple and unified framework that allows us to clarify the modification of fundamental radiative processes in NZI media of dimension $d$. Our formulation is convenient as it provides the necessary physical insight to understand how radiative transitions are affected by the material parameters and number of dimensions concomitantly.  This is relevant since some metamaterial implementations of NZI media often exhibit a reduced dimensionality \cite{Vesseur2013,Briggs2013,Li2016,Liberal2017photonic,Reshef2017direct,Luo2018coherent}.

We start by following the quantization procedure proposed by Milonni \cite{Milonni1995,Milonni2003} , where the two-level system can be modelled with the following Hamiltonian (See details in SM \cite{Supp1}):
\begin{equation}
\widehat{H}=\frac{\hslash \omega}{2}\widehat{\sigma}_z + \sum_{\mathbf{k},\lambda}\hslash\omega_k \widehat{a}^{\dagger}_{\mathbf{k}\lambda}\widehat{a}_{\mathbf{k}\lambda}
+ \sum_{\mathbf{k},\lambda}\hslash \left(g_{\mathbf{k}\lambda}\widehat{\sigma}^{\dagger}\widehat{a}_{\mathbf{k}\lambda}  + h.c.\right),
\label{eq:H}
\end{equation}
with $\widehat{\sigma}_z=\left|e\right\rangle\left\langle e\right|-\left|g\right\rangle\left\langle g\right|$, $\widehat{\sigma}^{\dagger}=\left|e\right\rangle\left\langle g\right|$  , $\omega$ being the transition frequency of the emitter and $\omega_k$  the eigenfrequency of the mode with wavevector $\mathbf{k}$. The sums run over all optical modes of wavevector $\mathbf{k}$, polarization $\lambda$ with unit polarization vector $\mathbf{e}_{\mathbf{k}\lambda}$, and annihilation operator $\widehat{a}_{\mathbf{k}\lambda}$. The coupling between the emitter and optical modes is characterized by the coupling strength
\begin{equation}
g_{\mathbf{k}\lambda}=-i\sqrt{\frac{Z\left(\omega_k\right)}{n_g\left(\omega_k\right)}}\sqrt{\frac{\hslash\omega_k}{2\varepsilon_0 V_d}}
\,\,\mathbf{p}\cdot\mathbf{e}_{\mathbf{k}\lambda}.
\label{eq:g_k}
\end{equation}

The impact of the background medium and its dispersion properties in the light-matter coupling are described by the presence of the normalized wave impedance $Z\left(\omega_k\right)$ and the group index $n_g\left(\omega_k\right)$ in Eq.\,(\ref{eq:g_k}). $V_d$ is the $d$-dimensional quantization volume.

Next, the relevant transition rates can be computed by using Fermi's golden rule \cite{Loudon2000}. For instance, the $A_{21}'$ coefficient corresponding to the rate of spontaneous emission reduces to
\begin{equation}
A_{21}'=2\pi\sum_{\mathbf{k}\lambda}\left|g_{\mathbf{k}\lambda}\right|^2\delta\left(\omega_k-\omega\right).
\label{eq:A21}
\end{equation}

This basic equation provides an often overlooked but important physical insight. It conveys the thought that the decay rate of a quantum emitter depends on the number of available optical modes, $\sum_{\mathbf{k}\lambda}$, and how strongly it couples to them $\left|g_{\mathbf{k}\lambda}\right|^2$ --- both factors must be taken into account in order to correctly describe the physics. Thinking in terms of how the modes are asymptotically depleted in a system (e.g., because the refractive index goes to zero) would not provide the complete physical picture if the coupling strength scales inverse proportionally in the zero-index limit. For this reason, it is in principle possible for the spontaneous emission rate to converge to zero, infinity or to a finite value in the zero-index limit.

To further emphasize this point, we rewrite Eq.\,(\ref{eq:A21}) as the product of two factors:  
$A_{21}'=G\left(\omega\right)N\left(\omega\right)$, describing (i) $G\left(\omega\right)$, how strongly the emitter couples to the optical modes as a function of the background, and (ii) $N\left(\omega\right)$, which describes the number of available modes. These factors are defined as follows:
\begin{align}
G\left(\omega\right)&=\left|\frac{g_{\mathbf{k}\lambda}}{g_{\mathbf{k}\lambda}^0}\right|^2 _{\omega_k \rightarrow \omega}
=\frac{Z\left(\omega\right)}{n_g\left(\omega\right)}
\label{eq:G}\\
N\left(\omega\right)&=
2\pi\sum_{\mathbf{k},\lambda}\left|g_{\mathbf{k}\lambda}^0\right|^2\delta\left(\omega_k-\omega\right)\nonumber\\
&=A_d\,\,\frac{\left|\mathbf{p}\right|^2}{\hslash\varepsilon_0}
\,\,\frac{\omega^d}{c^d}
\,\,\left|\mathrm{Re}\left[n(\omega)\right]\right|^{d-1}\,\,n_g\left(\omega\right),
\label{eq:N}
\end{align}

\noindent with $A_1=1/2$, $A_2=1/4$ and $A_3=1/(3\pi)$, and $g_{\mathbf{k}\lambda}^0$ is the vacuum limit of $g_{\mathbf{k}\lambda}$.

Understanding the explicit dependence on these two factors as a function of the material parameters and number of dimension provides a comprehensive picture on how different NZI media modify radiative processes. First, $N\left(\omega\right)$ is defined as the decay rate that would be observed if we could couple to existing modes in the dispersive medium, but with the coupling strength for modes in vacuum. Consequently, $N\left(\omega_Z\right)$ gives a good account on the modification of the number of modes induced by the material parameters. In particular, its dependence on the background is contained within the factor $n^{d-1}\left(\omega\right)\,\,n_g\left(\omega\right)$. This scaling rule can be understood since the sum over all modes is transformed into an integral $\int_0^{\infty}\,\,k^{d-1}\,dk$. In general, the number of modes depletes as the refractive index goes to zero, and this behavior is observed to be independent of the class of NZI media. $N\left(\omega\right)$ only depends on the refractive index, and the depletion in the NZI limit is stronger for a larger number of dimensions.  

A very different behavior is observed in terms of how strongly we couple to these modes. Specifically, $G\left(\omega\right)$ is defined as the magnitude square of the ratio between the coupling strength and its vacuum counterpart. Its scaling rule with respect to the background, given by $Z\left(\omega\right)/n_g\left(\omega\right)$, is \emph{independent} of the number of dimensions, but it critically depends on the class of NZI media. This behavior can be intuitively understood by noting that the interaction Hamiltonian is defined within the electric dipole approximation
$\widehat{H}_I=-\widehat{\mathbf{p}}\cdot\widehat{\mathbf{E}}$, and, therefore, $\left|g_{\mathbf{k}\lambda}\right|^2$ is proportional to the electric field intensity. Importantly, the background material modifies the strength of the electric field fluctuations per unit of energy, thus modifying the strength of how the modes couple to the emitter. Since the classical energy per mode can be written as
$U_{\mathbf{k}\lambda}=2\varepsilon_0 V\left|\mathbf{E}^{(+)}\right|^2/(Z\left(\omega_k\right)/n_g\left(\omega_k\right))$ (see SM \cite{Supp1}), it is clear that the electric field intensity per energy unit is modified by the factor $Z\left(\omega_k\right)/n_g\left(\omega_k\right)$ due to the material properties. In this manner, we find that materials with a high, or even diverging normalized wave impedance, like ENZ media, will tend to enhance radiative transitions compared to other classes of NZI media. 

Ultimately, it is the product between $G\left(\omega\right)$ and $N\left(\omega\right)$ that provides the total decay rate. By combining Eqs.\,(\ref{eq:G}) and (\ref{eq:N}) we obtain the compact expression:

\begin{equation}
A_{21}' = Z(\omega)\,\,\left|\mathrm{Re}\left[n(\omega)\right]\right|^{d-1}\,\,A_{21}.
\label{A21gen}
\end{equation}

By applying this general equation to the different NZI cases, i.e. at $\omega=\omega_Z$, one can note that the inhibition of spontaneous emission is not valid for all dimensions, even if the refractive index approaches zero. In fact, depending on the interplay between the normalized wave impedance and refractive index, one can observe either suppressed, finite, or even divergent decay rates in the NZI limit (Table 1 and Figs.\,\ref{fig:Fig1SpontEM3D}b-c for 1D and 2D cases).

The Purcell factor $A_{21}'/A_{21}$ might also take a constant value (2D ENZ or 1D EMNZ media) or present a divergent behavior (1D ENZ media), in accordance with previous studies in dispersive ENZ waveguides \cite{Alu2009boosting,Fleury2013,Sokhoyan2013,Sokhoyan2015,Li2016}. One might be tempted to justify this behavior as an example of Purcell enhancement in slow-light waveguides \cite{Lodahl2015}. However, this reasoning fails to explain all NZI cases. For instance, 1D MNZ is also a slow-light waveguide, with a near-zero group velocity at the MNZ frequency, and yet at this point, spontaneous emission is inhibited (see SM ection IV for a discussion on the validity of our theory to model implementations of 1D ENZ and MNZ media with dispersive waveguides \cite{Supp1}). 

\begin{table}[h]
    \centering
\begin{tabular}{|l|c|c|c|c|}
    \hline
      & $A'_{21}/A_{21}$ & ENZ & MNZ & EMNZ\\
    \hline
    1D & $Z(\omega_Z)$ & $\infty$ &$0$ & $\sqrt{\frac{\frac{d\mu(\omega)}{d\omega }\Bigr|_{\substack{\omega=\omega_Z}}}{\frac{d\varepsilon(\omega)}{d\omega }\Bigr|_{\substack{\omega=\omega_Z}}}}$\\
    \hline
    2D & $Z(\omega_Z)n(\omega_Z)$ & $|\mu(\omega_Z)|$ &$0$ &$0$ \\
    \hline
    3D & $Z(\omega_Z)n(\omega_Z)^2$ & 0 & 0 & 0 \\
    \hline
    
\end{tabular}
\caption{Purcell factor at $\omega_Z$ for ENZ, MNZ and EMNZ media in 1D, 2D and 3D.}
\label{Table}
\end{table}

The influence of the dimensionality on the absorption and stimulated emission rates can be obtained by repeating the same procedure used for spontaneous emission (see SM section III \cite{Supp1}). It confirms that these two rates are identical in all instances, and that the ratio between stimulated and spontaneous emission rates is given by the number of photons per optical mode. Therefore, we find that stimulated emission and absorption rates are always proportional to the spontaneous emission rate, as imposed by the very structure of the interaction Hamiltonian given by Eq.\,(\ref{eq:H}). It is then concluded that the ratios between the different radiative processes are fixed, and cannot be modified by changing the background medium.

In conclusion, we investigated dimension-dependent fundamental radiative processes in NZI media. Our formalism illustrates that in order to get the correct physical picture it is crucial to consider both the number of optical modes that may couple to an emitter as well as the coupling strength. These quantities are found to depend highly on the material class and the number of spatial dimensions. For example, we theoretically worked out a dimension-dependent Purcell factor leading to an inhibition of spontaneous emission in most NZI cases, but an enhanced Purcell factor inside 1D ENZ materials. Based on detailed-balance considerations, it can be readily found that other radiative processes such as stimulated emission and absorption follow the modifications induced on spontaneous emission. 

\section*{Acknowledgments}

R.W.B., N.E. and E.M acknowledge support from the Defense Advanced Research Projects Agency (DARPA) Defense Sciences Office (DSO) Nascent program and from the US Army Research Office. This work was performed while M.L. was a recipient of a Fellowship of the Belgian American Educational Foundation. M.L. and ENK would like to thank Daryl Vulis and Yang Li for stimulating discussion on zero-index topics.
OR acknowledges the support of the Banting Postdoctoral Fellowship of the Natural Sciences and Engineering Research Council of Canada (NSERC).ENK is supported by the National Science Foundation Graduate Research Fellowship Program under Grant No. DGE1144152 and DGE1745303. Any opinions, findings, and conclusions or recommendations expressed in this material are those of the author(s) and do not necessarily reflect the views of the National Science Foundation.

\bibliography{Main}

\end{document}